\pdfoutput=1
\PassOptionsToPackage{unicode}{hyperref}
\PassOptionsToPackage{hyphens}{url}
\documentclass[
]{book}
\usepackage{amsmath,amssymb}
\usepackage{lmodern}
\usepackage{iftex}
\ifPDFTeX
  \usepackage[T1]{fontenc}
  \usepackage[utf8]{inputenc}
  \usepackage{textcomp} 
\else 
  \usepackage{unicode-math}
  \defaultfontfeatures{Scale=MatchLowercase}
  \defaultfontfeatures[\rmfamily]{Ligatures=TeX,Scale=1}
\fi
\IfFileExists{upquote.sty}{\usepackage{upquote}}{}
\IfFileExists{microtype.sty}{
  \usepackage[]{microtype}
  \UseMicrotypeSet[protrusion]{basicmath} 
}{}
\makeatletter
\@ifundefined{KOMAClassName}{
  \IfFileExists{parskip.sty}{%
    \usepackage{parskip}
  }{
    \setlength{\parindent}{0pt}
    \setlength{\parskip}{6pt plus 2pt minus 1pt}}
}{
  \KOMAoptions{parskip=half}}
\makeatother
\usepackage{xcolor}
\usepackage{longtable,booktabs,array}
\usepackage{calc} 
\usepackage{etoolbox}
\makeatletter
\patchcmd\longtable{\par}{\if@noskipsec\mbox{}\fi\par}{}{}
\makeatother
\IfFileExists{footnotehyper.sty}{\usepackage{footnotehyper}}{\usepackage{footnote}}
\makesavenoteenv{longtable}
\usepackage{graphicx}
\makeatletter
\def\maxwidth{\ifdim\Gin@nat@width>\linewidth\linewidth\else\Gin@nat@width\fi}
\def\maxheight{\ifdim\Gin@nat@height>\textheight\textheight\else\Gin@nat@height\fi}
\makeatother
\setkeys{Gin}{width=\maxwidth,height=\maxheight,keepaspectratio}
\makeatletter
\def\fps@figure{htbp}
\makeatother
\providecommand{\tightlist}{%
  \setlength{\itemsep}{0pt}\setlength{\parskip}{0pt}}
\usepackage{booktabs}
\usepackage{multirow}
\usepackage{float}

\newcommand{\BigO}[1]{\ensuremath{\mathop{}\mathopen{}\mathcal{O}\mathopen{}\left(#1\right)}}

\usepackage{caption}

\ifLuaTeX
  \usepackage{selnolig}  
\fi
\usepackage[]{natbib}
\IfFileExists{bookmark.sty}{\usepackage{bookmark}}{\usepackage{hyperref}}
\IfFileExists{xurl.sty}{\usepackage{xurl}}{} 
\hypersetup{
  pdftitle={Simple Symmetric Sustainable Sorting --- the greeNsort® article},
  pdfauthor={Jens Oehlschlägel},
  hidelinks,
  pdfcreator={LaTeX via pandoc}}

\title{Simple Symmetric Sustainable Sorting \newline --- the greeNsort® article}
\author{Jens Oehlschlägel}
\date{2024-02-02}

\begin{document}
\maketitle

{
\setcounter{tocdepth}{1}
\tableofcontents
}
\newpage

\hypertarget{abstract}{%
\chapter{Abstract}\label{abstract}}

We explored an uncharted part of the solution space for sorting algorithms: the role of symmetry in divide\&conquer algorithms. We found/designed novel simple binary Quicksort and Mergesort algorithms operating in contiguous space which achieve improved trade-offs between worst-case CPU-efficiency, best-case adaptivity and RAM-requirements. The \emph{greeNsort®} algorithms need less hardware (RAM) and/or less energy (CPU) compared to the prior art. The new algorithms fit a theoretical framework: \emph{Footprint} KPIs allow to compare algorithms with different RAM-requirements, a new \emph{definition} of sorting API-targets simplifies construction of stable algorithms with mirrored scan directions, and our ordinal machine model encourages robust algorithms that minimize access \emph{distance}. Unlike earlier \emph{Quicksorts}, our \emph{Zacksort}, \emph{Zucksort} and \emph{Ducksort} algorithms optimally marry CPU-efficiency and tie-adaptivity. Unlike earlier \emph{Mergesorts} which required 100\% distant buffer, our \emph{Frogsort} and \emph{Geckosort} algorithms achieve similar CPU-efficiency with 50\% or less local buffer. Unlike natural Mergesorts such as \emph{Timsort} which are optimized for the best case of full-presorting, our \emph{Octosort} and \emph{Squidsort} algorithms achieve excellent bi-adaptivity to presorted best-cases without sacrificing worst-case efficiency in real sorting tasks. Our \emph{Walksort} and \emph{Jumpsort} have lower Footprint than the impressive low-memory \emph{Grailsort} and \emph{Sqrtsort} of Astrelin. Given the current climate-emergency, this is a call to action for all maintainers of sorting libraries, all software-engineers using custom sorting code, all professors teaching algorithms, all IT professionals designing programming languages, compilers and CPUs: check for better algorithms and consider symmetric code-mirroring.

\hypertarget{introduction}{%
\chapter{Introduction}\label{introduction}}

\citet{UKParliamentPOST:2022} reported a global energy consumption of 555 TWh for data centers and user devices in 2020 (excluding cryptomining, networks and TVs) and is projected to grow to 720 TWh in 2030.
IBM estimated that sorting - a basic component of software - is responsible for 25\% of computing (\citet{Mehlhorn:1977})\footnote{The first non-analog computer was a pure sorting machine designed by Herman Hollerith and sold by a predecessor of IBM in 1890}. If these old estimates hold, sorting costs 11.5 12TWh Nuclear Power Stations (NPS) in 2020 or 15 NPS in 2030, plus embodied energy resp. greenhouse gas for the production of data centers and user devices. Software developers usually consume sorting from few off-the-shelf sorting libraries, improving these libraries can massively save runtime energy (operational costs) and required hardware (embodied costs). Particularly reducing hardware requirements allows using older machines longer, which leads to more sustainable amortization of embodied costs. This paper reports the results of the \emph{greeNsort®} project (2010 - 2023), which developed simple sorting algorithms that need less runtime energy and/or less required hardware.

\newpage

\hypertarget{sustainability-measures}{%
\section{Sustainability measures}\label{sustainability-measures}}

The classic academic KPI for empirically comparing algorithms is runtime in seconds (\(runTime\)). However, runTime ignores embodied costs and does not allow fair comparison of different sorting algorithms because they require different amounts of memory relative to data size \[\%RAM = (dataRAM+bufferRAM)/dataRAM\] with an expected trade-off: lower memory costs (embodied) correlate with higher runtime costs (operational). A straightforward Footprint KPI combining embodied and operational costs is the integral of required hardware over runtime, i.e.~\[tFootprint = avg(\%RAM) * runTime\] Because operational energy (\(Energy\) measured at RAPL Socket) is highly correlated with runtime, similar reasoning leads to the energy footprint calculated as \[eFootprint = avg(\%RAM) * Energy\] Note that cloud \emph{Functions as a Service} (FaaS) such as AWS Lambda has the payment metric \emph{Memory over Time}, this is exactly \emph{tFootprint}.

\hypertarget{scope-of-greensort}{%
\section{\texorpdfstring{Scope of \emph{greeNsort®}}{Scope of greeNsort®}}\label{scope-of-greensort}}

\emph{greeNsort®} investigates binary divide\&conquer algorithms for general comparison in-memory sorting. This investigation and the design of new algorithms is guided by the following values and principles: \emph{greeNsort®} aims on algorithms that are \protect\hyperlink{sustainability}{\emph{sustainable}}, \protect\hyperlink{generality}{\emph{general}}, directly \protect\hyperlink{stability}{\emph{stable}}, \protect\hyperlink{robustness}{\emph{robust}}, \protect\hyperlink{resilience}{\emph{resilient}}, \protect\hyperlink{scalability}{\emph{scalable}}, \protect\hyperlink{reliability}{\emph{reliable}}, \protect\hyperlink{adaptivity}{\emph{adaptive}}, potentially \protect\hyperlink{concurrency}{\emph{parallel}}, \protect\hyperlink{simplicity}{\emph{simple}} and \protect\hyperlink{beauty}{\emph{beautiful}}.

\hypertarget{values-and-principles}{%
\chapter{Values and principles}\label{values-and-principles}}

\hypertarget{sustainability}{%
\section{Sustainability}\label{sustainability}}

The overarching goal of \emph{greeNsort®} is providing algorithms that are sustainable in the sense of minimizing \(eFootprint\) and that are suitable to replace less sustainable algorithms in many places. Hence the following values guided the development:

\hypertarget{generality}{%
\section{Generality}\label{generality}}

The new algorithms should be applicable in as many contexts as possible. This implies comparison sorting on any data types, including the possibility to sort elements of varying size.

\hypertarget{stability}{%
\section{Stability}\label{stability}}

The new algorithms should directly support stable sorting

\hypertarget{robustness}{%
\section{Robustness}\label{robustness}}

The new algorithms should perform robustly on different types of hardware, whatever the particular features regarding random access, cache-size, branch prediction etc., hence comparisons are done between algorithms not tuned to specific hardware. This implies that \emph{greeNsort®} does not assume a specific machine model like a random-access model with constant costs for single-element access or a disk-model with constant costs for access to blocks of data. It is simple assumed, that there is a monotonic relation between access distance and access-cost.

\hypertarget{resilience}{%
\section{Resilience}\label{resilience}}

The algorithmic portfolio should include algorithms that perform in extreme situations and on old hardware with little (RAM) resources.

\hypertarget{scalability}{%
\section{Scalability}\label{scalability}}

The new algorithms should reliably perform \(N \log{N}\) in the worst-case, hence only divide\&conquer algorithms are considered.

\hypertarget{reliability}{%
\section{Reliability}\label{reliability}}

The new algorithms should reliably perform \(N \log{N}\) or better for any patterns of input data. Deterministic algorithms have advantages.

\hypertarget{adaptivity}{%
\section{Adaptivity}\label{adaptivity}}

The new algorithms should be adaptive, but not tuned to best cases. The main task of sorting algorithms is to sort data that is not already sorted, hence algorithms are developed to perform well in worst (or average) cases.

\hypertarget{concurrency}{%
\section{Concurrency}\label{concurrency}}

The new algorithms should allow proper and easy parallelization, hence the algorithms are described and demonstrated as top-down divide\&conquer, not bottom-up. Experts can still implement without recursion.

\hypertarget{simplicity}{%
\section{Simplicity}\label{simplicity}}

The new algorithms should be easy to adopt, i.e.~write, to debug and to verify. Hence the new algorithms must be simple. This is almost the most important single value, except for the following:

\hypertarget{beauty}{%
\section{Beauty}\label{beauty}}

The new algorithms should be beautiful. Computer scientists might be surprised that an aesthetic value is listed here. Note: \emph{Simplicity} is an aesthetic value. Spoiler: \emph{Symmetry} is \emph{the} single most important ingredient of the new sustainable algorithms.

\hypertarget{symmetry}{%
\chapter{Symmetry}\label{symmetry}}

In many disciplines symmetry plays an important role, particularly bi-symmetry. Examples are vertebrates in biology, vehicles in engineering and churches in architecture. By contrast, in writing program code symmetry seems to play no relevant role! The von-Neumann-Machine is rife with asymmetries:

\begin{itemize}
\tightlist
\item
  \textbf{Access-asymmetry} the fact that memory access is asymmetric either the left element first then the right one or the right element first and then the left one
\item
  \textbf{Buffer-asymmetry} the fact that buffer placement relative to data is asymmetric, data may either be placed left of buffer memory (DB) or right of buffer memory (BD)
\end{itemize}

as is the topic of sorting

\begin{itemize}
\tightlist
\item
  \textbf{Order-asymmetry} the fact that that `order' is asymmetric and reaches from `low' to `high' `keys'
\item
  \textbf{Pivot-asymmetry} the fact that a binary pivot-comparison (one of LT, LE, GT, GE) assigns an element equal to the pivot either to one partition or the other
\item
  \textbf{Tie-asymmetry} the fact that stable ties are asymmetric, they may represent their original order either from left to right (LR) or from right to left (RL)
\end{itemize}

Bi-Symmetry as a design principle is rare, and if, bi-symmetry is mostly used on a loop-level. The secret ingredient of the new \emph{greeNsort®} algorithms is the \emph{symmetric-asymmetry} principle often found in nature, engineering and arts: design low-level asymmetry and turn it into high-level symmetry by \emph{mirroring}. For algorithmic-code this means: do not fight but embrace loop-level asymmetries and create recursion-level symmetry. Instead of the usual divide\&conquer self-recursive functions, \emph{greeNsort®} uses mutual-recursive code-mirrored functions. Code-mirroring enhances the algorithmic solution space, which contains many unexplored areas and surprising solutions. This begins with a new definition of `sorting':

\hypertarget{definition-of-sorting}{%
\chapter{Definition of sorting}\label{definition-of-sorting}}

According to Knuth \citet{Knuth:1998}, p.~1 sorting is \emph{``the rearrangement of items into ascending or descending order''} and can be distinguished into \emph{stable} and \emph{unstable} sorting. This definition creates four different goals for sorting ``\emph{unstable ascending}'', ``\emph{unstable descending}'', ``\emph{stable ascending}'', and ``\emph{stable descending}''. Sorting libraries implement all four or a subset of these four. From Knuth's mathematical perspective the definition of sorting is perfect.

\begin{figure}
\centering
\includegraphics{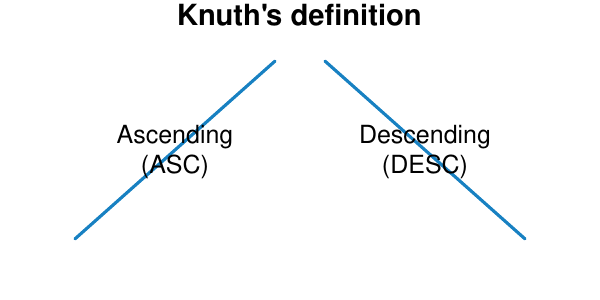}
\caption{\label{fig:unnamed-chunk-1}Knuth's definition of sorting}
\end{figure}

However, in the context of computers, algorithms are not abstract, they operate on \emph{elements} that are stored in \emph{memory} that is addressed from left to to right address \emph{locations} (address locations are notated here from left to right in order to not confuse this with low and high sorting keys). Habitually ascending and descending sequences are written from \emph{left} to \emph{right} :

\begin{figure}
\centering
\includegraphics{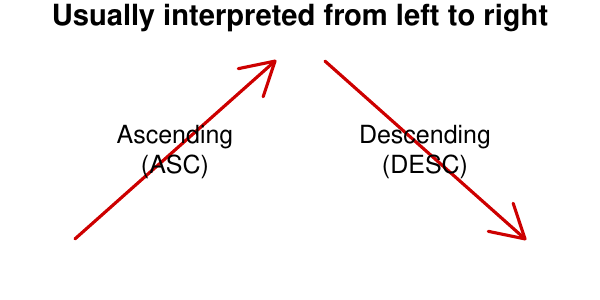}
\caption{\label{fig:unnamed-chunk-2}Conventional interpretation: from left to right}
\end{figure}

The two abstract sorting sequences \emph{asc} and \emph{desc} correspond to four concrete sorting sequences in memory: \emph{ascleft}, \emph{ascright}, and \emph{descleft}, and \emph{descleft}. The Difference between \emph{descleft} and \emph{ascleft}, is a reverted - but stable - sequence of ties!\\

\begin{figure}
\centering
\includegraphics{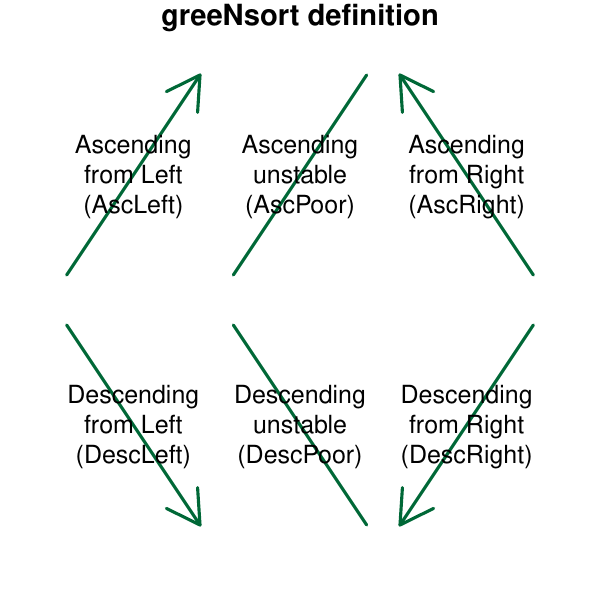}
\caption{\label{fig:unnamed-chunk-3}greeNsort definition: unstable sorting has two API targets (AscPoor, DescPoor) but stable symmetric sorting has four API targets (AscLeft, AscRight, DescLeft and DescRight)}
\end{figure}

The \emph{greeNsort}® definition is powerful because it facilitates reasoning in an increased solution space; there is similarity to the invention of the \emph{imaginary part} which increased the number space from \emph{real numbers} to \emph{complex numbers}, such that suddenly the square-root of a negative number was defined.

\hypertarget{merging-or-partitioning}{%
\chapter{Merging or partitioning?}\label{merging-or-partitioning}}

Divide\&Conquer sorting falls into two categories: Split\&Merge (most work in the merge phases) or Partition\&Pool (most work during partitioning phases). Which is better? A prototypical stable deterministic balanced binary Mergesort uses 100\% buffer and uses per recursion-level one read-scan with comparisons for the input-data and one write-scan for the merge. A prototypical stable deterministic balanced binary Partitionsort uses 100\% buffer, one read-scan and further writing and reading for calculating an (approximate) median as pivot, a read-an-compare-scan for counting the size of the resulting (two) partitions and finally a read-scan and a write-scan for the actual partitioning. Certain sacrifices allow to save some of this work: dropping reliability (determinism) saves the work for calculating the median, further dropping stability and generality allows to saves the buffer and the counting-scan and leads to an attractive family of algorithms: Quicksorts.

\hypertarget{quicksort-dilemma}{%
\chapter{Quicksort dilemma}\label{quicksort-dilemma}}

Instead of writing stable partitions with the help of a buffer, the Quicksort family of algorithms is characterized by SWAPing pairs of elements in-place, hence sacrificing\footnote{Note that SWAPing directly equally-sized elements directly sacrifices generality, indirect sorting by SWAPing pointers to variable-sized elements causes random-access and sacrifices robustness.} stability. Quicksort is tricky, wrong implementations easily lead to quadratic runtime, particularly if deterministic pivots are used, or even leads to endless loops, particularly if sentinels are used. Note that its inventor Tony Hoare defined Quicksort as a probabilistic algorithm using random pivots. Many historic attempts to tune Quicksort with deterministic pivots have finally been regretted because it made Quicksort vulnerable for unexpected runtime and algorithmic complexity attacks. In the following we assume random pivots.

Compared to inefficient approaches like the Lomuto-scheme, the brilliant idea of Hoare in \emph{Quicksort1} was to write the two partitions outside-in from the left and the right, this saves a counting-scan and promised to isolate pivot-ties in a third partition in the middle \citetext{\citeyear{Hoare:1961a}; \citealp{Hoare:1961b}}. Hoare's symmetric design obviously aimed on efficient scaling (average \(\BigO{N\log{N}}\) operations) and early-terminating on ties (\(\BigO{N\log{D}}\), where \(D\) is the number of distinct keys). However, on certain data inputs \emph{Quicksort1} degenerated to \(\BigO{N^2}\).

Several attempts have been made to avoid quadratic runtime for arbitrary data inputs. \citet{Sedgewick:1977a} compared Hoare's algorithm with two different approaches: asymmetric partitioning and a symmetric partitioning going back to \citet{Singleton:1969}, where the two pointers from the left and right both stop on pivot-ties before SWAPing. This guarantees a balanced partitioning (and hence average \(\BigO{N\log{N}}\)), but it SWAPs ties and it does not early terminate on ties (hence not \(\BigO{N\log{D}}\)). Sedgewick concluded that the asymmetric approach was vulnerable to quadratic runtime and recommended the symmetric version: \emph{Quicksort2} became the standard for many years.

In an attempt to fix quadratic runtimes in \emph{Quicksort2} implementations with deterministic pivots and in order to gain early-termination on ties, \citet{Wegner:1985} and later \citet{Bentley:1993} developed \emph{Quicksort3}\footnote{we report a simpler version that is slightly faster in the worst-case of untied keys and slightly slower in the best case of tied keys, see also \citet{vonLeitner:2007}} that collects pivot-ties in a third partition in the middle between the low and high partition. \emph{Quicksort3} achieves \(\BigO{N\log{D}}\) for tied data, but due to extra-operations for identifying and placing the ties it is slower than \emph{Quicksort2} for untied data. This trade-off between worst-case efficiency and best-case efficiency we term the `Quicksort dilemma'.

\citet{Yaroslavskiy:2009} achieved a notable improvement with dual-pivot Quicksort (\emph{Dupisort}\footnote{our slightly better \emph{Tricksort} removes a dead-branch from Yaroslavskiy's code}) that uses its extra-operations to create a real third partition. This is faster and no longer a binary sort: it improves the average \(\BigO{N\log_2{N}}\) algorithm to \(\BigO{N\log_3{N}}\) regarding moves, but the algorithm is strongly serial and difficult to implement branchless. Then, \citet{Edelkamp:2016a} published an even faster and simpler binary Block-Quicksort that reduced branch-mispredictions. However, Block-Quicksort had only a rudimentary and expensive early-termination mechanism, that we skipped in \emph{Quicksort2B}. At this point, history returned in a huge cycle\footnote{Indeed a cycle: the optimization published by Edelkamp\&Weiß (2016) was already suggested in a little known paper of \citet{Hoare:1962} in which he gives context and explanations for his 1961 algorithm} to \emph{Quicksort2}, still leaving the quicksort dilemma unresolved.

Stepping back and analyzing the commonalities of all those attempts, it seems that all authors assume that partitioning must by symmetric, and for early-termination on ties a third partition is necessary. However, the \emph{greeNsort®} analysis of the problem finds that

\begin{itemize}
\tightlist
\item
  for early-termination not three, not two but one partition is sufficient, the algorithm just needs to be able to diagnose an all-tied partition and stop recursing deeper into this branch
\item
  in order to get pure partitions where ties are not in multiple partitions, the partitioning must be Mutually Exclusive and Collectively Exhaustive (MECE), this requires to make a clear (asymmetric) decision to which side pivot-ties go
\item
  it is possible to provably optimally modify an asymmetric partitioning for detection of all-tied data (DIET-Method)
\item
  it is possible to make asymmetric partitioning reliable against any adverse input pattern (FLIP-method)
\end{itemize}

\hypertarget{diet-method}{%
\section{DIET-method}\label{diet-method}}

Distinct Identification for Early Termination (DIET) works as follows: in the main loop of asymmetric binary partitioning, when searching for a pair of elements that require SWAPing, one pointer stops at pivot-ties, the other does not stop on pivot-ties. Assume the algorithm begins to search with the pointer that stops on ties. DIET adds a pre-loop that searches from the same side for a non-pivot-tied-key. If this pre-loop reaches the other side without finding a non-tie, all data are tied and the algorithm can exit recursion. If the pre-loop finds a non-tie, almost no work is lost and can be reused: only the last (non-tie) elements needs to be compared a second time against the pivot to find out to which side it belongs. I.e. we just set the pointer back one element and enter the main loop. That's it. And it provably costs exactly one extra comparison (and one extra pointer increments and decrement) per partitioning, hence less than \(N\) extra comparisons. That is the unavoidable price of tuning for early-termination on ties. If we use this DIET-partitioning in the usual self-recursive manner, we obtain \emph{Zocksort}, which still is vulnerable for certain (asymmetric) inputs. Now we reached a similar point like Sedgewick, but we do not try to fix these beautifully efficient loops (or give up). Instead we FLIP.

\hypertarget{flip-method}{%
\section{FLIP-method}\label{flip-method}}

Fast Loops In Partitioning (FLIP) means to embrace the fast loops of the asymmetric partitioning and to create symmetry on the recursion level by left-right-mirroring. An input pattern that fools an asymmetric partitioning must itself be asymmetric, for example a \emph{Zocksort} that partitions pivot-ties to the lower partition is fooled by data with two distinct keys, many high, and one low: random pivots are mostly high, hence all data goes to the left partition and the algorithm does not progress. Note that this data input cannot fool a Zocksort that partitions pivot-ties to the higher partition. By recursively alternating between an asymmetric partitioning and its left-right-mirrored twin, no asymmetric pattern can fool the resulting algorithm. Zig-zagging between the two twins is called \emph{Zacksort} (`Zack' is the German word for `zag' and implies a connotation of `quick'). An elaboration is the \emph{Zucksort} algorithm which flips the partitioning asymmetry only in the recursive branch that can contain the pivot-ties, in the other branch it does not matter (`Zuck' is the German word for `twitch' hence also implies a connotation of `quick').

\hypertarget{poet-method}{%
\section{POET-method}\label{poet-method}}

Pre-Order Early Termination (POET) is a tuning replacement of the DIET pre-loop in order to early-terminate on presorted data: instead of looping while equaling the pivot, the algorithm loops while the data does not violate the desired sorting order. If the loop reaches the other side the algorithm can early terminate, if it detects an out-of-order element, it resets the pointer to the starting point and enters the main loop. Note that Early-Termination on ties is a special case of POET. Note that POET unlike DIET is not provably optimal for early-termination on ties because the work of the pre-loop cannot be reused. The resulting algorithm is called \emph{Ducksort}.

\hypertarget{partial-sorting}{%
\section{Partial sorting}\label{partial-sorting}}

Several variants of partial sorting are easily derived from \emph{Quicksort}. A quite generic version (\emph{Quickpart}) takes two parameters position \texttt{l} and position \texttt{r} in sorted order and guarantees sorted order between \texttt{l} and \texttt{r}. Other definitions of partial sorting are special cases thereof. For example the C++ \texttt{std::partial\_sort} guarantees the smallest elements until position \texttt{middle} to be sorted, this is achieved by setting \texttt{l=middle} and \texttt{r=max} position. Specifying \texttt{l==r} gives the popular \emph{Quickselect} published by Hoare as \emph{FIND} \citetext{\citeyear{Hoare:1961a}; \citealp{Hoare:1961c}}. Note that \emph{Quickselect} still does partial sorting work in order to guarantee that the l==r-th element is in the l==r-th position and that all elements smaller than the selected element are left of \texttt{l} and that all elements greater than the selected element are right of \texttt{r}. Note further that \emph{Quickselect} does not define which element it selects in case of ties. But the well-defined tie-handling of \emph{Zackselect}, \emph{Zuckselect} or \emph{Duckselect} returns the leftmost and rightmost positions of elements tied with the element at the desired position, for some tricky details see the code.

\hypertarget{branch-prediction}{%
\section{Branch-prediction}\label{branch-prediction}}

Like Block-Quicksort from \citet{Edelkamp:2016a} these \emph{greeNsort®} algorithms such as \emph{Ducksort}, \emph{Duckpart}, \emph{Duckselect} can be tuned to behave branchless, hence faster (named \emph{DucksortB}, \emph{DuckpartB}, \emph{DuckselectB}). Note that \emph{Pdqsort} of Orson Peters \citetext{\citeyear{Peters:2014}; \citealp{Peters:2015}; \citealp{Peters:2021a}; \citealp{Peters:2021b}} is a related asymmetric algorithm that when necessary involves a mirrored asymmetric partitioning. It does not achieve high-level symmetry and is not provably optimal, for example uses some heuristic shuffling. However Peters excellent C++ implementation is branchless with tuning for ties and tuning for presorted data and little overhead and it has been formally proven correct {[}Lammich:2020{]}, hence for all practical purposes I can highly recommend it as mostly faster than the \emph{greeNsort®} research implementations. However, once combined with an expensive comparison function such as localized string comparison \texttt{strcoll}, \emph{Pdqsort} becomes slower than \emph{DucksortB}. Note also that \emph{Pdqsort} is implemented with deterministic pivots and fallback to potentially slower \emph{Heapsort}, and note that the \emph{greeNsort®} algorithms can be implemented like this as well, without any need for heuristic shuffling. Finally note that the simple branch-parallel \emph{PDucksortB} outperforms \emph{Pdqsort}, which has a serial dependence as the author states himself.\footnote{``I don't have plans currently. I would have to do some research on modern standard C++ parallel programming, and there are some tricky things in PDQsort if you want to parallelize. In particular it is assumed the left partition is recursed on first.'' \url{https://github.com/orlp/pdqsort/issues/16\#issuecomment-823145493}}

\hypertarget{quicksort-conclusion}{%
\section{Quicksort conclusion}\label{quicksort-conclusion}}

Embracing low-level asymmetry and addressing high-level bi-symmetry elegantly solves the decade-old Quicksort-dilemma in an easily provable way, which suggests applying symmetric recursion to other algorithms as well: to overcome the limitations of \emph{Quicksorts}. While \emph{Quicksorts} are memory parsimonious and have good cache-locality, they lack stability, deterministic worse-case guarantees and the ability to sort elements of varying size, and they are difficult to parallelize. Hence let's turn to \emph{Mergesorts}.

\hypertarget{mergesort-trilemma}{%
\chapter{Mergesort trilemma}\label{mergesort-trilemma}}

In the Quicksort-dilemma there were trade-offs between efficiency and adaptivity. In merging there are trade-offs between efficiency, adaptivity and reducing buffer memory: one cannot have all three of them: full \(\BigO{N\log{N}}\) efficiency and full \(\BigO{N}\) adaptivity and low-memory.

\hypertarget{forgotten-art}{%
\section{Forgotten art}\label{forgotten-art}}

Let's rehearse how a generic efficient potentially parallel \emph{Mergesort} is organized. Naive implementations allocate 100\% buffer in the merge, copy the data to the buffer (100\% moves), merge the two input streams back (100\% comparisons and 100\% moves) and deallocate the buffer. Obviously better is to allocate the buffer once before recursing and de-allocating once done, however what about the 200\% moves?
\emph{Timsort} (\citet{Peters:2002}) allocates not more than 50\% buffer, copies note more than 50\% out and merges 100\% back, that is 150\% moves.
\citet{Sedgewick:1990} teached a \emph{Nocopy-Mergesort} that only merges (100\% moves) without copying out by alternating between two memory regions of size 100\% each.

Sedgewick teaches \emph{Nocopy-Mergesort} with three loop-checks: two on the two input-streams and one on the output loop, obviously two loop-checks on the two input-streams ar enough. Sedgewick explains how to get away with one loop-check on the output streams by using sentinels in a mutual-recursive \emph{Bitonic-Mergesort}, which sorts the left half ascending and the right half descending (from left), which is however not stable. Using the symmetric sorting definition we can fix this, by sorting the right half not descending from left but ascending from right which makes the algorithm nicely symmetric (\emph{Bimesort}). However there are still disadvantages: this sentinel-approach must move data in each merge, hence cannot skip all merging for presorted data. Interestingly this sentinel-approach is not even necessary to get down to one loop-check: \citet{Knuth:1973} had teached 20 years earlier that one loop-check is sufficient, if only that input-sequence is checked for exhaustion from which the last element was merged. Combining this with \emph{Nocopy-Mergesort} gives an efficient algorithm we name \emph{Knuthsort} in honor of Donald Knuth. Note that \citet{Katajainen:1997} showed that the number of loop-checks can be further reduced to ½ by investigating which of the two input-sequences exhausts first and only checking that one. Be warned that Katajainen's code reads beyond the last element, we fixed this and named the result \emph{Katasort}. Note further, that Katajainen's optimization costs an extra comparison, hence we consider it tuned and prefer \emph{Knuthsort} as the prior-art reference for efficient prior-art binary \emph{Mergesort} in contiguous space.

For random data \emph{Knuthsort} (and \emph{Katasort}) are much more efficient than \emph{Timsort}. Conventional wisdom assumes, that Timsort's inefficient merge is needed to reduce buffer to 50\% and to achieve \(\BigO{N}\) adaptivity to presorted data: after the merge \emph{Timsort} has the data in the same memory-region as before the merge, hence it is easy to skip copying-out and merging-back in case of presorted keys. However, conventional wisdom is wrong, we can have one of the two without compromising on efficiency, either buffer-reduction or full adaptivity. Let's start with the latter.

\hypertarget{full-adaptivity}{%
\section{Full adaptivity}\label{full-adaptivity}}

The alternating merge of \emph{Nocopy-Mergesort} finally completes in the data memory, not in the buffer. However, for odd recursion depths this starts merging from the buffer memory, hence Sedgewick copies the data to the buffer memory before recursion. Our \emph{Omitsort} no longer uses a pre-determined alternating merge, instead we let the merge functions decide whether it merges (or whether it can omit the merge in case of presorted data): the merge function simply returns the location of the data and the sort function checks whether after its two merges the data are in the same memory region, if not, it copies one to the data region. For fully presorted data no moves are needed anymore, not even Sedgewicks initial moves. \emph{Omitsort} uses \(\BigO{N}\) comparisons for diagnosing presortedness as non-overlap of the two input sequences.

For descending data, an ascending \emph{Omitsort} cannot omit: each merge must do its work, hence the total cost are \(\BigO{N\log{N}}\). Cheaper would be to do nothing and only reverse the total sequence after checking presortedness throughout the recursion. Our \emph{Octosort} achieves this by no longer requiring a specific order during the merging: the resulting order is simply data-driven and returned by the recursive sort function. Only if the left and right recursive sorts return different directions, the desired (ascending) order is enforced. For perfectly ascending and descending data the are no moves during the recursion, for descending data only a final reversal is needed, hence \emph{Octosort} is like \emph{Timsort} \(\BigO{N}\) bi-adaptive for pre-sorted data, but unlike \emph{Timsort} much more efficient for non-sorted data. \emph{Octosort} can also be fully parallelized, while \emph{Timsort} has inherently serial tasks.

\hypertarget{distance-minimization}{%
\section{Distance minimization}\label{distance-minimization}}

\begin{quote}
``For decades, the machine balance of compute speed to memory bandwidth has increased 15\%--30\% per year {[}\ldots{]} Projections for future machines only exacerbate the current data movement crisis'' \citep{Dongarra:2022}.
\end{quote}

We all have learned that sorting cost is \(\BigO{N\log{N}}\) \ldots{} for constant access costs in the RAM-model. We all know that the RAM-model is wrong and access costs are not constant: todays memory hierachies are deep, L1, L2, L3, RAM on local socket, RAM on foreign socket, and todays virtualization and cloud techniques might even give us memory on different machines, in different LAN and across WAN networks.

The traditional answers to the data movement crisis are k-ary algorithms (which reduce the number of memory passes), block-access (which is only suitable for one of the cache-layers) and cache-oblivious algorithms (which are complicated). \emph{greeNsort®} takes a different perspective, the perspective of an \emph{Ordinal Machine Model}: there is a cost for moving data over distance, but the exact cost of a move (or access) over distance \(d\) is unknown. It could be anything between \(\BigO{1}\), \(\BigO{\log{d}}\) and \(\BigO{d}\). An example for the latter - linear move costs - would be sorting N cars in a row in a parking slot next to a road. Let's assume another free N positions of buffer parking space. A buffer next to the N cars is the best we can hope for, hence merging N cars to the buffer space costs moving each car on average N positions on each recursion level, which gives us total move cost of \(\BigO{N^2\log_2{N}}\) for all cars on all recursion levels. That's expensive.

Note that even \emph{K-ary Mergesort} for our cars still would be quadratic: \(\BigO{N^2\log_k{N}}\). Contrast this to \emph{Quicksort}, where the move distance is halved on each partitioning in the recursion, therefore the total move costs is \(\BigO{N^2}\). This is a little appreciated reason why \emph{Quicksort} performs robustly on many different machines. In other words, \emph{Quicksort} zooms into local memory, while \emph{Mergesort} keeps merging over global distances.

\hypertarget{gapped-merging}{%
\section{Gapped-merging}\label{gapped-merging}}

A simple trick brings locality to \emph{Mergesort}: for \(N\) elements of data take \(2N\) elements of memory and alternate merging between odd and even positions, this reduces the move distances as the algorithm divides deeper. Unfortunately gapped-merging comes with a couple of disadvantages: reading \(N\) elements actually scans \(2N\), hence gapped \emph{GKnuthsort} is on current CPUs slower than \emph{Knuthsort}. Also gapped-merging requires all elements to have the same size.

\hypertarget{buffer-merging}{%
\section{Buffer-merging}\label{buffer-merging}}

Locality requires starting the merging with local buffer gaps between the data. If we want a contiguous result of the sort with only two regions, data and buffer, this requires that we merge not only the data but also the buffer. Rehearse this: buffer becomes a first-class object of merging. Understand that Divide\&Conquer \emph{with} buffer and good locality requires Divide\&Conquer \emph{of} buffer. Once this is understood, it looks really weird, that in \emph{Mergesort} for decades we only merged data but not buffer. Now let's look at some ways to merge buffer. Let uppercase letters represent data and lowercase represent buffer.

If the result of buffer merging is one contiguous data region and one contiguous buffer region, we have to make a fundamental asymmetric choice: data left and buffer right (\texttt{Ab}) or data right and buffer left (\texttt{aB}). If we assume \texttt{Ab} before splitting and after merging, standard self-recursion implies to replicate that pattern across all recursion levels, i.e.~we merge from \texttt{AbCd} to \texttt{ACbd}. In order to merge \texttt{A} and \texttt{C}, a naive approach would first transport all input streams to the right: \texttt{bdAC} and then merge back to \texttt{ACbd}. 100\% extra moves are expensive (\emph{TKnuthsort}). A more efficient approach would only relocate the streams in the left half to the gaps in the right half \texttt{dbCA} before merging to the left \texttt{ACbd}. Note that the order of the streams has changed, hence care is needed to retain stability when merging (\emph{Crocosort} with Knuth's merge). Unfortunately, with 50\% extra moves, this is still as inefficient as \emph{Timsort}.

\hypertarget{symmetric-merging}{%
\section{Symmetric merging}\label{symmetric-merging}}

It is possible to get rid of any extra moves if we leverage symmetric mutual recursion: if in the left branch we sort the data to the left (buffer to the right) and in the right branch we sort the data to the right (buffer left). What we obtain is the data in the outer regions and the buffer in the inner regions: the two inner buffer regions are contiguous without any extra moves. That's nice, but even nicer is that not more than 50\% buffer is needed: merging is done from inner to outer input-streams such that the result is aligned at the outer border. This semi-inplace merging has the following adaptivity properties: for perfectly presorted data the number of comparisons and moves is automatically reduced to 50\%, with a non-overlap test the number of comparisons can be reduced to 0\%, but at lat least 50\% of the data must be moved. Symmetric merging offers two variants: a symmetric variant that sorts one branch left-ascending and one branch right-ascending (\emph{Geckosort}) and a asymmetric variant which uses the same order in both branches (\emph{Frogsort}). \emph{Geckosort} is 25\% adaptive to ascending and 25\% adaptive to descending keys. \emph{Frogsort} is 50\% adaptive to presorting in the implemented order. Note that \emph{FROG} is an acronym of \emph{Flip Recursive Organized Gaps}.\footnote{Gaps are crucial for stable sorting. As late as 2006, \citet{Bender:2006} published \emph{Librarysort}, a Gapped Insertion Sort. Like people leave gaps between books in bookshelves, gaps reduce insertion costs from \(\BigO{N^2}\) to amortized \(\BigO{N\log{N}}\) in the RAM-model. While the usage of gaps in \emph{Librarysort} is rather probabilistic, \emph{Frogsort} uses gaps in a deterministic, `Organized' way.}

\hypertarget{frogsort0}{%
\section{Frogsort0}\label{frogsort0}}

The simple (and first) variant of \emph{Frogsort} was \emph{Frogsort0} (2010): it operates on \emph{triplets} of memory elements, one buffer element symmetrically in the middle between two data elements, i.e.~\texttt{AbC}. \emph{Frogsort0} splits and merges the triplets. Rule: if there is an odd number of triplets, the surplus triplet goes to the outer side. The setup of the gapped elements can be done before Split\&Merge.

\begin{table}[H]

\caption{\label{tab:Frogsort0}Sketch of Frogsort0}
\centering
\begin{tabular}[t]{r|c|c|c|c|c|c|c|c|c}
\hline
position & 1 & 2 & 3 & 4 & 5 & 6 & 7 & 8 & 9\\
\hline
setup & A & b & C & D & e & F & G & h & I\\
\hline
merge & A & C & b & e & D & F &  &  & \\
\hline
merge & A & C & D & F & b & e & h & G & I\\
\hline
merge & A & C & D & F & G & I & b & e & h\\
\hline
\end{tabular}
\end{table}

Odd number of elements can be handled by using one extreme dummy value (see the implementation of \emph{Frogsort0}) or by a specific third recursion function that handles the outermost incomplete triplet (see the implementation of \emph{PFrogsort0}).

\hypertarget{frogsort1}{%
\section{Frogsort1}\label{frogsort1}}

An alternative is splitting single elements in \emph{Frogsort1}. Rule 1: if there is an odd number of elements, the surplus element goes to the outer side. Rule 2: the size of the buffer is \(N/2\) (integer division). The setup is done in an extra recursion before the Split\&Merge Recursion.

\begin{table}[H]

\caption{\label{tab:Frogsort1}Sketch of Frogsort0}
\centering
\begin{tabular}[t]{r|c|c|c|c}
\hline
position & 1 & 2 & 3 & 4\\
\hline
top-2 & A & b & C & \\
\hline
top-1 & A & C & b & D\\
\hline
top & A & C & D & b\\
\hline
\end{tabular}
\end{table}

\hypertarget{frogsort2}{%
\section{Frogsort2}\label{frogsort2}}

Textbook knowledge tells us that Divide\&Conquer should be balanced in order to minimize operations. Symmetric merging allows to reduce the buffer to much less than 50\%: let the inner branch be \(p\%\), then symmetric merging needs only \(p\%\) buffer. Yes, this increases the maximum recursion depth and the total number of operations, but it reduces the total \%RAM. Hence for the sustainable Footprint measure, we expect an U-shaped function of \(p\). \emph{greeNsort®} implemented this as \emph{Frogsort2}. For sorting doubles, surprisingly, not only exists an optimal \(p\) below 10\% regarding Footprint, there is also an optimal \(p\) below 20\% regarding speed. This is surprising given usual space-speed trade-off expectations, that less RAM implies longer runTime. One reason for this is: imbalanced merging makes for better branch-prediction. With \emph{Frogsort2} we have a nice algorithm, that can be tuned to specific hardware features using its parameter \(p\). Note that \emph{Frogsort1} is a special case of \emph{Frogsort2} at \(p=0.5\).

\hypertarget{frogsort3-and-6}{%
\section{Frogsort3 and 6}\label{frogsort3-and-6}}

So far we have \emph{split} the buffer between the two branches. An alternative is to \emph{share} the buffer, i.e.~first we send all the buffer down the left branch, and then we send the buffer down the right branch. This allows to use even less buffer. \emph{Frogsort3} does this, until there is enough buffer at a branch to switch to \emph{Frogsort1}. \emph{Frogsort6} does this, until there is enough buffer at a branch to switch to \emph{Frogsort2}. \emph{Frogsort1} is a special case of \emph{Frogsort3} with parameter \(p=0.5\). \emph{Frogsort6} has two parameters \(p3\) and \(p2\), and \emph{Frogsorts} 1,2,3 are special cases of \emph{Frogsort6}.

\hypertarget{squidsort}{%
\section{Squidsort}\label{squidsort}}

\emph{Frogsort} saves 50\% compares and moves for presorted data, but not for reverse-sorted data. \emph{Geckosort} saves 25\% compares and moves for presorted data and for reverse-sorted data. Tuning \emph{Frogsort} with a non-overlap comparison for presorted data reduces all other compares to 0\%. Combining \emph{Frogsort} with the data-driven lazily enforced order of \emph{Octosort} gives \emph{Squidsort}, which needs 0\% comparisons and 50\% moves for presorted and reverse-sorted data (see \emph{Squidsort1} and \emph{Squidsort2}). \emph{Squidsort} beats \emph{Timsort} and related algorithms such as \emph{Peeksort} and \emph{Powersort} by \citet{Munro:2018}, unless data is extremely presorted.

\hypertarget{mergesort-conclusion}{%
\chapter{Mergesort Conclusion}\label{mergesort-conclusion}}

Embracing low-level asymmetry and addressing high-level bi-symmetry elegantly improves the trade-offs of the decade-old Mergesort-trilemma. The symmetric definition of sorting guides the development of stable algorithms, and exploiting buffer-asymmetry in symmetric-merging (and stable symmetric partitioning) allows to reduce the amount of buffer required, in some cases even using less energy and sorting faster (\emph{Frogsort2}).

\hypertarget{further-algorithms}{%
\chapter{Further algorithms}\label{further-algorithms}}

\hypertarget{low-memory-merging}{%
\section{Low-memory merging}\label{low-memory-merging}}

In-place merging algorithms have enjoyed huge academic attention. Several in-place Mergesorts are known for decades but were too slow in practice. In 2008 (or earlier) Sedgewick created the term ``holy sorting grail'' for a practically usable stable \(\BigO{[N\log{N}}\) in-place-Mergesort. In 2013 \citet{Astrelin:2013} published \emph{Grailsort} an implementation of \citet{Huang:1992} that was not dramatically slower than other \emph{Mergesorts}, hence deserved its name. Since 2014 Sedgewick redefined the ``holy sorting grail'' and required a \(\BigO{N}\) best case. Anyhow, in-place merging is overrated, Astrelin also published \emph{Sqrtsort} which is faster than \emph{Grailsort} and needs only a practically negligible \(\BigO{\sqrt{N}}\) buffer. The \emph{greeNsort®} algorithm \emph{Walksort} also requires \(\BigO{\sqrt{N}}\) buffer, is faster, and its equally fast variant \emph{Jumpsort} uses relocation-moves for distance minimization (like \emph{Crocosort}).

Regarding speed and energy for sorting random data, these \(\BigO{\sqrt{N}}\) buffer algorithms are inferior to \emph{Frogsort} and \emph{Squidsort}. Regarding Footprint the ranking for sorting random doubles is \emph{Frogsort2} \textless{} \emph{Squidsort2} \textless{} \emph{Frogsort3} \textless{} \emph{Walksort} \textless{} \emph{Jumpsort} \textless{} \emph{Frogsort1} \textless{} \emph{Squidsort1} \textless{} \emph{Sqrtsort} \textless{} \emph{Grailsort} \textless{} \emph{Knuthsort}. For pre-sorted data the ranking of \emph{Walksort} and \emph{Jumpsort} improves: they are quite adaptive.

\hypertarget{partitionpool}{%
\section{Partition\&Pool}\label{partitionpool}}

Some of the \emph{greeNsort®} learnings can be transferred from \emph{Split\&Merge} to \emph{Partition\&Pool} algorithms. It is possible to turn \emph{Zacksort} into the stable \emph{Kiwisort} algorithm using 100\% distant buffer. By partitioning not only data but also 100\% local buffer we get the distance reducing \emph{Swansort}. For a very special use case it is even possible to reduce to 50\% buffer, see \emph{Storksort}.

\hypertarget{size-varying}{%
\section{Size-varying}\label{size-varying}}

A popular method for sorting elements of varying size - such as strings, bigints or arbitrary objects with a key field - is indirect sorting of pointers to elements using \emph{Quicksort}, or better one of \emph{Zacksort}, \emph{Zucksort} or \emph{Ducksort}. However, this is inefficient because it incurs heavy random access (see \emph{UZacksort}, and even more inefficient if the sort is stabilized by breaking ties by comparing pointer address values, see \emph{WQuicksort2}). An alternative is directly sorting size varying elements using a buffer. Splitting can be done either by helper-pointers to elements or directly splitting at null-terminators (for the latter see \emph{VKnuthsort} and \emph{VFrogsort1}). These direct sorts are more efficient than the indirect Quicksorts. Only if there are many string duplicates, unstable \emph{UZacksort} benefits from its \(\BigO{N\log{D}}\) early termination. Hence it seems promising to implement direct stable Partition\&Pool algorithms for size-varying data (\emph{VKiwisort} and \emph{VSwansort}, not yet implemented).

\hypertarget{parallel}{%
\section{Parallel}\label{parallel}}

Doing a task in parallel can save energy, Intel calls this ``race to idle''. All Divide\&Conquer algorithms are easily implemented with parallel branches (see \emph{PQuicksort2}, \emph{PQuicksort2B}, \emph{PDucksort}, \emph{PDucksortB}). However, the partitioning of the \emph{Quicksort} family is difficult to implement in parallel, and this involves trade-offs. By contrast, parallelizing binary merges is relatively straightforward. \emph{greeNsort®} uses a method that allows to parallelize over an arbitrary number of processes, not only power of two processes, see \emph{PKnuthsort}, \emph{PFrogsort0}, \emph{PFrogsort1}, \emph{PFrogsort2}, \emph{PFrogsort3}, \emph{PVKnuthsort}, \emph{PVFrogsort1}. As expected, parallel execution reduces not only runTime but also Energy (although to a lesser extent).

Note that the parallel speed-up of the \emph{Frogsorts} scales almost as linear with the number of cores like the speed-up of \emph{Knuthsort}, hence the Footprints of \emph{PFrogsorts} are clearly lower than that of \emph{PKnuthsort}.

Note that the benefit of \emph{Frogsort2} over \emph{Frogsort1} somewhat diminishes when more parallel cores are used.

Note that the linear setup-phase of \emph{Frogsort0} somewhat better parallelizes than the recursive setup of \emph{Frogsorts1,2,3,6}. Hence it is promising to implement variants of \emph{Frogsorts2,3,6} which leverage a linear setup with chunks of a predefined mixture of data and buffer elements.

\hypertarget{results}{%
\chapter{Results}\label{results}}

\hypertarget{methods-and-measurement}{%
\section{Methods and measurement}\label{methods-and-measurement}}

For energy measurement we use the RAPL counters of the linux \texttt{powercap} kernel module, see \texttt{lib\_energy.h} and \texttt{lib\_energy.c}. All measurements reported here are done on an Intel i7-7700 CPU under ubuntu.20.04 with the 5.13.0-44-generic kernel and compiling our testbed with gcc.9.4.0. The CPU is run with hyper-threading switched of in the bios. The algorithms are measured on the following input data patterns:

\begin{itemize}
\tightlist
\item
  \texttt{permut}: randomly permuted numbers from \(1 \dots n\)
\item
  \texttt{tielog2}: random sample of \(\log_2{n}\) distinct values
\item
  \texttt{ascall}: \(n\) distinct ascending numbers
\item
  \texttt{asclocal}: \(n\) distinct numbers randomly put into \(\sqrt{n}\) presorted sequences of length \(\sqrt{n}\)
\item
  \texttt{ascglobal}: \(n\) distinct numbers cut into ascending \(\sqrt{n}\) quantiles of length \(\sqrt{n}\) and randomly permuted per quantile
\end{itemize}

Measurements for these 5 patterns are averaged to a \texttt{TOTAL} KPI for ranking of algorithms. Furthermore the following 3 patterns are measured

\begin{itemize}
\tightlist
\item
  \texttt{descall}: \(n\) distinct descending numbers
\item
  \texttt{desclocal}: \(n\) distinct numbers randomly put into \(\sqrt{n}\) reverse-sorted sequences of length \(\sqrt{n}\)
\item
  \texttt{descglobal}: \(n\) distinct numbers cut into descending \(\sqrt{n}\) quantiles of length \(\sqrt{n}\) and randomly permuted per quantile
\end{itemize}

The descending patterns are interesting with regard of the symmetry of adaptivity, but not included in the \texttt{TOTAL} KPI. We believe that adaptivity to ascending data is more important than to descending, furthermore, including them would give too much weight to easy patterns.

\hypertarget{quicksort-results}{%
\section{Quicksort results}\label{quicksort-results}}

\begin{figure}[H]
\includegraphics[width=1\linewidth]{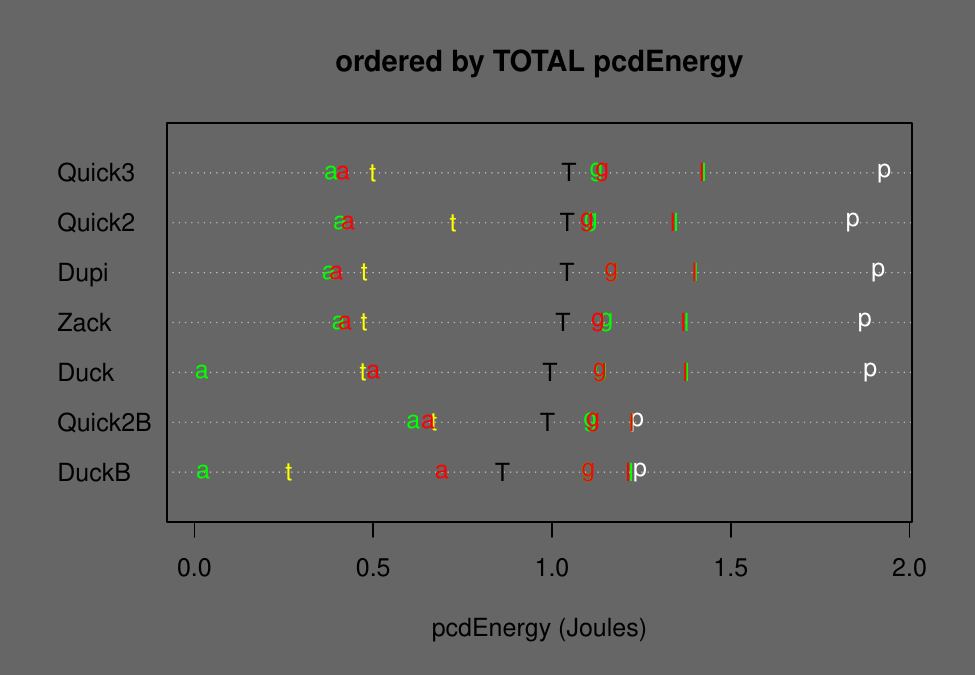} \caption{Medians of Quicksort alternatives ordered by TOTAL Energy. T=TOTAL, p=permut, t=tielog2; green: a=ascall, g=ascglobal, l=asclocal; red:  a=descall, g=descglobal, l=desclocal}\label{fig:Quickallplot}
\end{figure}

\begin{table}[H]

\caption{\label{tab:Ducktab2}Ducksort / Quicksort2 (ratios of medians and p-values from two-sided Wilcoxon tests)}
\centering
\begin{tabular}[t]{lrrrrrr}
\toprule
  & r(\%M) & r(rT) & r(pcdE) & d(pcdE) & p(rT) & p(pcdE)\\
\midrule
TOTAL & 1 & 0.95 & 0.95 & -0.05 & 0 & 0\\
\cmidrule{1-7}
ascall & 1 & 0.05 & 0.05 & -0.39 & 0 & 0\\
\cmidrule{1-7}
descall & 1 & 1.17 & 1.16 & 0.07 & 0 & 0\\
\cmidrule{1-7}
ascglobal & 1 & 1.02 & 1.02 & 0.03 & 0 & 0\\
\cmidrule{1-7}
descglobal & 1 & 1.02 & 1.03 & 0.04 & 0 & 0\\
\cmidrule{1-7}
asclocal & 1 & 1.02 & 1.02 & 0.03 & 0 & 0\\
\cmidrule{1-7}
desclocal & 1 & 1.02 & 1.03 & 0.03 & 0 & 0\\
\cmidrule{1-7}
tielog2 & 1 & 0.66 & 0.65 & -0.25 & 0 & 0\\
\cmidrule{1-7}
permut & 1 & 1.02 & 1.03 & 0.05 & 0 & 0\\
\bottomrule
\end{tabular}
\end{table}

\begin{table}[H]

\caption{\label{tab:Ducktab3}Ducksort / Quicksort3 (ratios of medians and p-values from two-sided Wilcoxon tests)}
\centering
\begin{tabular}[t]{lrrrrrr}
\toprule
  & r(\%M) & r(rT) & r(pcdE) & d(pcdE) & p(rT) & p(pcdE)\\
\midrule
TOTAL & 1 & 0.96 & 0.95 & -0.05 & 0 & 0.0000\\
\cmidrule{1-7}
ascall & 1 & 0.05 & 0.05 & -0.36 & 0 & 0.0000\\
\cmidrule{1-7}
descall & 1 & 1.20 & 1.20 & 0.08 & 0 & 0.0000\\
\cmidrule{1-7}
ascglobal & 1 & 1.02 & 1.01 & 0.01 & 0 & 0.0992\\
\cmidrule{1-7}
descglobal & 1 & 1.00 & 0.99 & -0.01 & 0 & 0.0160\\
\cmidrule{1-7}
asclocal & 1 & 0.97 & 0.97 & -0.05 & 0 & 0.0000\\
\cmidrule{1-7}
desclocal & 1 & 0.98 & 0.97 & -0.05 & 0 & 0.0000\\
\cmidrule{1-7}
tielog2 & 1 & 1.02 & 0.95 & -0.03 & 0 & 0.0000\\
\cmidrule{1-7}
permut & 1 & 0.98 & 0.98 & -0.04 & 0 & 0.0005\\
\bottomrule
\end{tabular}
\end{table}

\hypertarget{mergesort-results}{%
\section{Mergesort results}\label{mergesort-results}}

\begin{figure}[H]
\includegraphics[width=1\linewidth]{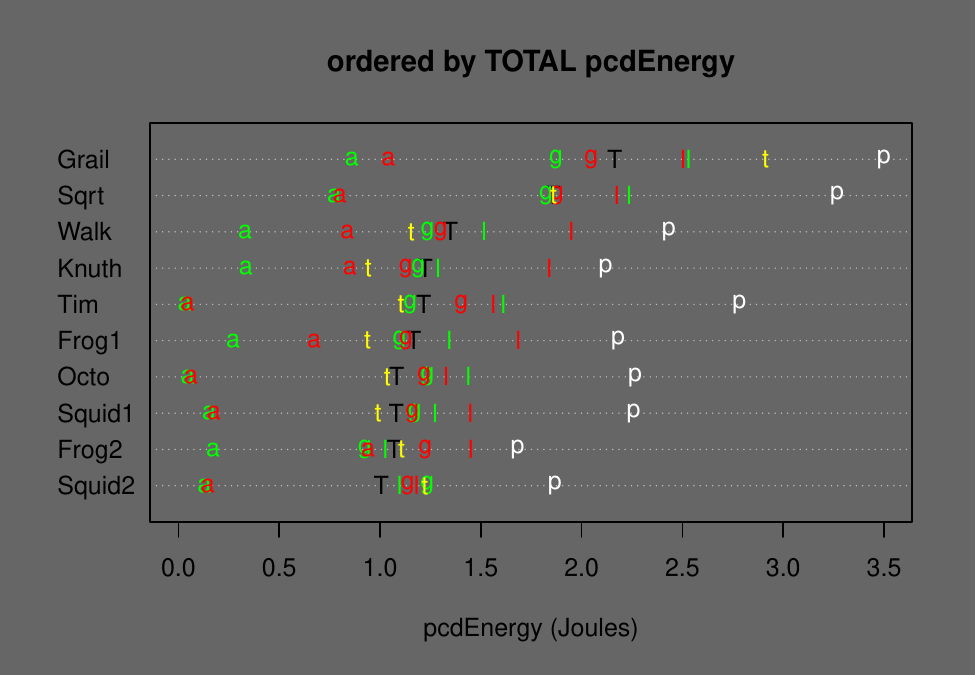} \caption{Medians of Mergesort alternatives ordered by TOTAL Energy. T=TOTAL, p=permut, t=tielog2; green: a=ascall, g=ascglobal, l=asclocal; red:  a=descall, g=descglobal, l=desclocal}\label{fig:MergeEnergyplot}
\end{figure}

\begin{figure}[H]
\includegraphics[width=1\linewidth]{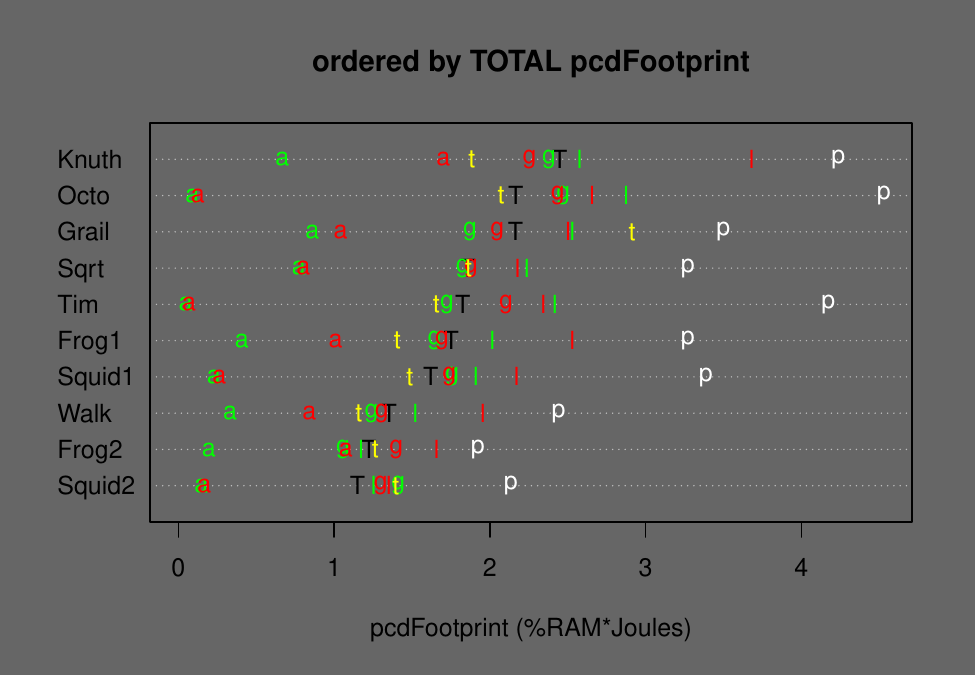} \caption{Medians of Mergesort alternatives ordered by TOTAL eFootprint. T=TOTAL, p=permut, t=tielog2; green: a=ascall, g=ascglobal, l=asclocal; red:  a=descall, g=descglobal, l=desclocal}\label{fig:MergeFootplot}
\end{figure}

\begin{table}[H]

\caption{\label{tab:Squidtab}Squidsort2 / Knuthsort (ratios of medians and p-values from two-sided Wilcoxon tests)}
\centering
\begin{tabular}[t]{l|r|r|r|r|r|r|r|r}
\hline
  & r(\%M) & r(rT) & r(pcdE) & r(pcdF) & d(pcdE) & p(rT) & p(pcdE) & p(pcdF)\\
\hline
TOTAL & 0.57 & 0.81 & 0.82 & 0.47 & -0.22 & 0.0000 & 0.0000 & 0\\
\cline{1-9}
ascall & 0.57 & 0.38 & 0.38 & 0.22 & -0.21 & 0.0000 & 0.0000 & 0\\
\cline{1-9}
descall & 0.57 & 0.15 & 0.17 & 0.10 & -0.70 & 0.0000 & 0.0000 & 0\\
\cline{1-9}
ascglobal & 0.57 & 1.04 & 1.04 & 0.59 & 0.04 & 0.3595 & 0.4566 & 0\\
\cline{1-9}
descglobal & 0.57 & 1.00 & 1.01 & 0.58 & 0.01 & 0.1750 & 0.0023 & 0\\
\cline{1-9}
asclocal & 0.57 & 0.84 & 0.85 & 0.49 & -0.19 & 0.0000 & 0.0000 & 0\\
\cline{1-9}
desclocal & 0.57 & 0.64 & 0.64 & 0.37 & -0.66 & 0.0000 & 0.0000 & 0\\
\cline{1-9}
tielog2 & 0.57 & 1.30 & 1.30 & 0.74 & 0.28 & 0.0000 & 0.0000 & 0\\
\cline{1-9}
permut & 0.57 & 0.87 & 0.88 & 0.50 & -0.25 & 0.0000 & 0.0000 & 0\\
\hline
\end{tabular}
\end{table}

\hypertarget{related-work}{%
\chapter{Related work}\label{related-work}}

K-ary sorting algorithms can dramatically reduce movements compared to binary algorithms. Radix sorting algorithms go further and also reduce comparison cost, but are only available for certain data types and collation orders. \citet{Axtmann:2017} have developed the k-ary \emph{In-place Parallel Super Scalar Samplesort (IPS4o)} with the motivation to provide a more efficient alternative to quicksort variants in standard libraries. Their recent study improved and compared it to other algorithms including radix sorts and shows that \emph{IPS4o} outperforms binary and other k-ary algorithms for sufficiently large data sets \citep{Axtmann:2022}. Also their radix algorithms outperformed the quit generic library of \citet{Skarupke:2016}. This C++ library of the KIT is highly recommended. Some limitations of \emph{IPS4o} are its code complexity, that it is not stable and its limited adaptivity to presorted data.

\hypertarget{future-work}{%
\chapter{Future work}\label{future-work}}

\hypertarget{algorithms}{%
\section{Algorithms}\label{algorithms}}

We expect that it is possible to make \emph{IPS4o} stable using the permutation logic of our \emph{Walksort}. We suspect, \emph{Squidsort} could be made even more adaptive by replacing the static splits by searching for a optimal splitting point as in \emph{Peeksort}. For elements of varying size that have no easily searchable separator like the null-terminator in strings, versions of \emph{Frogsorts} should be developed which use pointers to elements (for direct sorting). Finally, we believe that parallel \emph{Frogsorts} 2,3,6 can be written using predefined chunks with a fixed number of data and buffer elements. Beyond sorting, we would not be surprised, if the symmetric recursion and particularly Divide\&Conquer with buffer could be used to design other algorithms, e.g.~clustering.

\hypertarget{sorting-apis}{%
\section{Sorting APIs}\label{sorting-apis}}

Current sorting APIs limit the efficiency of sorting, for example C's library (Quicksort) expects a ternary comparison function (using two binary comparisons under the hood), where an API expecting a binary comparison function would allow to sort with less comparisons. Optimal sorting APIs is an under-explored topic.

\hypertarget{code-mirroring}{%
\section{Code-Mirroring}\label{code-mirroring}}

In the testbed we have mirrored code sections by hand. In order to reduce the manual work of code-mirroring and to make errors less likely, code-mirroring could be done in meta-programming or programming languages accompanied with IDEs that support code-mirroring. In order to reduce binary code size, we envision code-mirroring CPU-instructions and compiler support. This is a promising novel field of research.

\hypertarget{call-to-action}{%
\chapter{Call to action}\label{call-to-action}}

Despite interesting open research questions, the currently available portfolio of algorithms - particularly from KIT and \emph{greeNsort®} - is rich enough to provide more sustainable alternatives for older established algorithms. In times of climate crisis it is imperative to reduce the energy and hardware consumption now. Hence we call all maintainers of sorting code, in languages, libraries and software to review their sorting-requirements and replace their algorithms with more sustainable ones. For further guidance turn to \href{https://greensort.org/}{\emph{greensort.org}} and \href{https://github.com/greensort/}{\emph{github.com/greensort}}.

This call to action is not limited to researchers and software-engineers, we also call all teaching computer scientists to update their didactic materials on algorithms and sorting such as textbooks, MOOCs etc. The power of the \emph{greeNsort®} methods and insights should be available to future generations of computer scientists without restriction from the outset.

\hypertarget{conclusion}{%
\chapter{Conclusion}\label{conclusion}}

\emph{greeNsort®} presents a new perspective on binary divide\&conquer sorting: \emph{Simple Symmetric Sustainable Sorting} deliberately departs from the state of the art and previous habits, enabling new algorithms. Our \emph{Footprint} KPIs combine traditional operational costs (time, energy) with hardware investment requirements (memory) and enable ranking of algorithms with different memory requirements. Our symmetric definition of sorting (four stable API targets instead of two) clarifies how to write robust algorithms. Our ordinal machine model promotes the development of robust algorithms by focusing on minimizing distances. New algorithms are enabled in an enlarged solution-space by embracing asymmetry (of partitioning and of loops in von-Neumann machines) and delegating symmetry into (mutual) recursion design. Our main results are are new Quicksort which solves the \emph{Quicksort-Dilemma} and a new Mergesort which resolves the \emph{Mergesort-Trilemma} (and novel variations of those algorithms).

60 years after Hoare's \emph{Quicksort}, we present the algorithm that Hoare wanted to invent: a symmetric probabilistic algorithm that early terminates on ties without compromising efficiency. \emph{Zacksort} resp. \emph{Zucksort} combine the symmetric \emph{FLIP} method with the lean \emph{DIET} method, which is provably optimal. It is possible to integrate adaptivity on presorted data: \emph{Ducksort} uses the \emph{POET} instead of the \emph{DIET} method: good in practice but no longer optimal.

75 years after John von Neumann's \emph{Mergesort}, we present stable algorithms that combine the buffer-locality of \emph{Quicksort} with the generality of \emph{Mergesort}. Instead of 100\% distant buffer, \emph{Frogsort} and \emph{Geckosort} need only 50\% (or less) local buffer. Furthermore they are automatically half-adaptive to presorted data, and we show how to achieve bi-adaptivity to pre-sorted and reverse-sorted data: \emph{Octosort} and \emph{Squidsorts} embrace existing order and only lazily enforce the desired order. Unlike natural Mergesorts such as \emph{Timsort} which are optimized for the best case of full-presorting, the \emph{greeNsort®} algorithms are optimized for the worst case of real sorting work and can be parallelized to multiple CPU-cores. They also can be implemented for size-varying elements (unlike Quicksort, which incurs expensive random-access when used indirectly). Our low-memory algorithms \emph{Walksort} and distance-reducing \emph{Jumpsort} have lower Footprint than the impressive \emph{Grailsort} and \emph{Sqrtsort} of Astrelin, and hence allow resilient use of old hardware with little RAM.

\hypertarget{author-project}{%
\chapter{Author \& Project}\label{author-project}}

Jens Oehlschlägel is a psychologist and Ph.D.~statistician who has programmed computers since 1977. He has worked in different roles and industries, including 10 years as Data Scientist at McKinsey. Jens has done innovative work in psychological testing (\href{https://www.testzentrale.de/shop/frankfurter-aufmerksamkeits-inventar-2.html}{\emph{Frankfurter Aufmerksamkeits-Inventar}}), statistical software (\href{https://cran.r-project.org/web/packages/bit}{\emph{bit}}, \href{https://cran.r-project.org/web/packages/bit64}{\emph{bit64}}, \href{https://cran.r-project.org/web/packages/ff}{\emph{ff}}) and cluster-analysis meta-algorithms (\href{http://www.truecluster.com/}{\emph{truecluster.com}}).
Since 2011 Jens is a member of the ACM, works as a database architect and devotes his spare time to the \emph{greeNsort®} project.

For more details, illustrations and measurement results see the \emph{greeNsort®} Innovation-Report and Code at \href{https://github.com/greeNsort/}{\emph{github.com/greeNsort}}. For further information consult \href{https://greensort.org/}{\emph{greensort.org}}. For news please follow us on twitter at \href{https://x.com/greeNsort_algos}{\emph{@greeNsort\_algos}}, or if twitter goes bankrupt, on mastodon at \href{https://scicomm.xyz/@greeNsort}{@greeNsort@scicomm.xyz} .

\emph{greeNsort®} is a protected trademark and may not be used without permission. For free open-source software there will be a self-certification program, that will allow those projects to promote their use of \emph{greeNsort®} algorithms. Beyond the self-certification program, for consulting or certification with the \emph{greeNsort®} brand or logo, please contact us by mail at \emph{consulting{[}at{]}greensort.eu} or \emph{certification{[}at{]}greensort.eu} .

\bibliography{greensort.article.bib,packages.bib}
\addcontentsline{toc}{chapter}{\bibname}

\end{document}